\documentclass[10pt,conference]{IEEEtran}
\IEEEoverridecommandlockouts
\usepackage{amsmath,amsfonts}
\usepackage{algorithmic}
\usepackage{graphicx}
\usepackage{textcomp}
\usepackage{xcolor}
\usepackage{tcolorbox}
\usepackage{graphicx}
\usepackage{listings}
\usepackage{makecell}
\usepackage{adjustbox}
\usepackage{multirow}
\usepackage{graphicx}
\usepackage{svg}
\usepackage{xspace}
\usepackage{amsmath}
\usepackage{subfigure}
\usepackage{caption}
\usepackage{amsmath}
\usepackage{booktabs}
\usepackage{color, colortbl}
\usepackage{url}
\usepackage{caption} 
\usepackage{pifont}

\newcommand{\xmark}{\ding{55}}
\newcommand{\cmark}{\ding{51}}

\DeclareUnicodeCharacter{2212}{-}
\definecolor{Gray}{gray}{0.9}
\definecolor{green}{RGB}{102,252,102}
\definecolor{ored}{RGB}{255,99,71}
\definecolor{orange}{RGB}{255,165,0}
\definecolor{lightgray}{RGB}{211,211,211}

\usepackage[ruled,vlined,linesnumbered]{algorithm2e}

\SetCommentSty{mycommfont}

\usepackage{tikz}
\usetikzlibrary{shapes}

\newcommand{\tool}{{\textsf{AdaT}}\xspace}

\newcommand{\sidong}[1]{\textcolor{blue}{\textbf{Sidong}: #1}}

\newcommand{\eq}[1]{\vspace{-10pt}{\footnotesize {#1} \normalsize}}

\definecolor{amber}{rgb}{1.0, 0.49, 0.0}
\definecolor{amberdark}{rgb}{1.0, 0.75, 0.0}

\def\BibTeX{{\rm B\kern-.05em{\sc i\kern-.025em b}\kern-.08em
    T\kern-.1667em\lower.7ex\hbox{E}\kern-.125emX}}
\begin{document}

\title{Efficiency Matters: Speeding Up Automated Testing with GUI Rendering Inference}

\author{\IEEEauthorblockN{Sidong Feng}
\IEEEauthorblockA{
\textit{Monash University}\\
Melbourne, Australia \\
sidong.feng@monash.edu}
\and
\IEEEauthorblockN{Mulong Xie}
\IEEEauthorblockA{
\textit{Australian National University}\\
Canberra, Australia \\
mulong.xie@anu.edu.au}
\and
\IEEEauthorblockN{Chunyang Chen\IEEEauthorrefmark{1}}
\IEEEauthorblockA{
\textit{Monash University}\\
Melbourne, Australia \\
chunyang.chen@monash.edu} \\
}

\maketitle

\begin{abstract}
% While the app under testing is mostly idle, the tool has to wait until the GUI finishes rendering before moving to the next event.
Due to the importance of Android app quality assurance, many automated GUI testing tools have been developed. Although the test algorithms have been improved, the impact of GUI rendering has been overlooked. On the one hand, setting a long waiting time to execute events on fully rendered GUIs slows down the testing process. On the other hand, setting a short waiting time will cause the events to execute on partially rendered GUIs, which negatively affects the testing effectiveness. An optimal waiting time should strike a balance between effectiveness and efficiency. We propose \tool, a lightweight image-based approach to dynamically adjust the inter-event time based on GUI rendering state. Given the real-time streaming on the GUI, \tool presents a deep learning model to infer the rendering state, and synchronizes with the testing tool to schedule the next event when the GUI is fully rendered. The evaluations demonstrate the accuracy, efficiency, and effectiveness of our approach. We also integrate our approach with the existing automated testing tool to demonstrate the usefulness of \tool in covering more activities and executing more events on fully rendered GUIs.
% The evaluations demonstrate our approach can achieve 99.8\% inference accuracy of discriminating rendering state and can detect 89\% of the bugs in a less run-time.
% We also integrate our approach with the existing automated testing tool to evaluate the usefulness of our \tool.
% Our experiments show that our enhanced testing tool can achieve 6.96\% higher activity coverage and execute 13.24\% more events on 88.81\% fully rendered GUIs.
\end{abstract}

\begin{IEEEkeywords}
Efficient android GUI testing, GUI rendering, Machine Learning
\end{IEEEkeywords}

\section{Introduction}

\begin{figure*}
	\centering
	\includegraphics[width=0.85\linewidth]{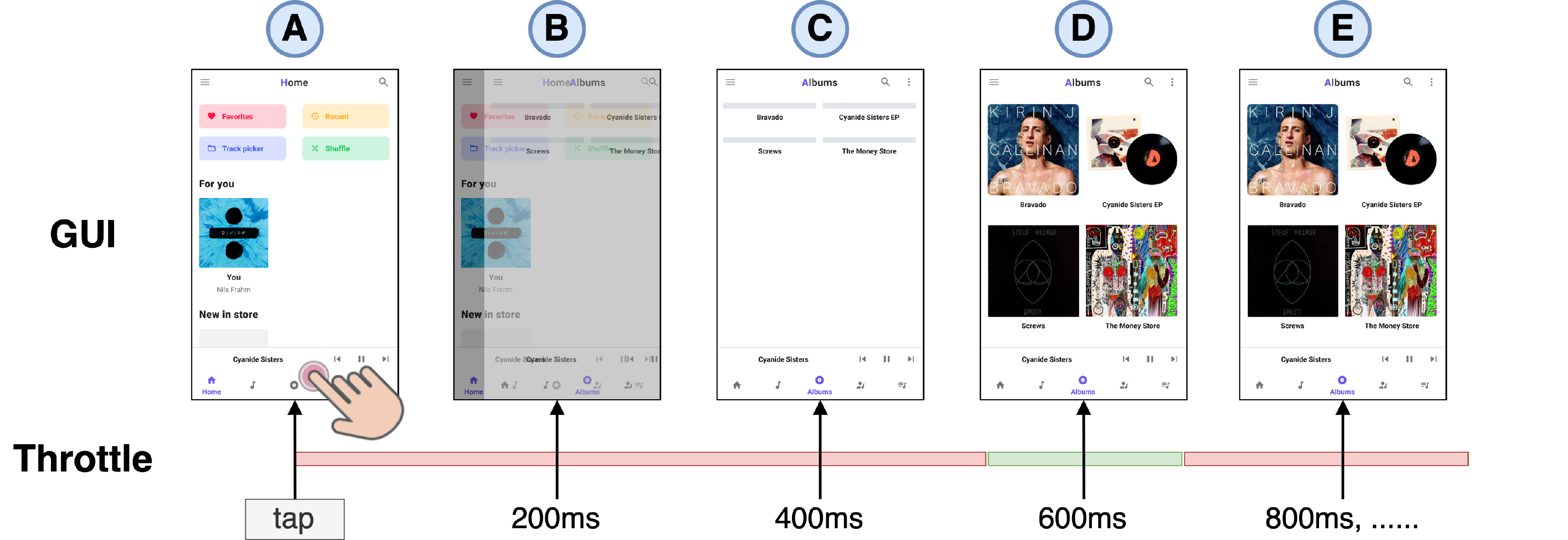}
	\caption{Automated GUI testing with different throttles. Green bars represent ideal throttle, red bars represent flawed throttles that ineffectively test on partially rendered states or inefficiently stagnate on GUI.}
	\label{fig:timeline}
% 	\vspace{-0.5cm}
\end{figure*}

\begin{comment}
Mobile applications have gained great popularity in recent years.
There have been over 3.8 million Android apps and 2 million iPhone apps striving to gain users on Google Play and Apple App Store~\cite{web:appstat}.
Bug-prone apps can significantly impact user experience and lead to a downgrade of their ratings, harming the reputation of app developers and companies~\cite{wang2018android}.
Therefore, to ensure quality, apps are extensively tested before they are released on the market.
While manual and scripted testing is a common practice, it is often laborious and time-consuming.
The ever-growing complexity of apps calls for automated testing solutions.
\end{comment}

GUI (Graphical User Interface) is one of the most common forms of user interface that provides a visual bridge between a software application and end-users through which they can interact with each other.
Since most bugs or issues can be spotted by users in GUI, GUI testing is widely used to ensure app quality.
There are many automated GUI testing works 
%Existing works have been focusing on designing sophisticated GUI exploration algorithms 
based on randomness~\cite{web:monkey,mao2016sapienz}, app artifact (e.g., source code, activity)~\cite{azim2013targeted,yang2013grey}, reverse engineering~\cite{gu2019practical,su2017guided}, and deep learning~\cite{li2019humanoid,degott2019learning}.
%to achieve high testing coverage, resulting in effective bug detection.
For most dynamic GUI testing tools, the more time for testing, the higher testing coverage, the more likely to find bugs, and the higher quality of the app release.
However, due to the budget limit and market pressure, development teams have to meet deadlines by striking a balance between testing time and other demands~\cite{joorabchi2013real}.
%Although testing coverage can be improved with sophisticated algorithms, one often overlooked aspect is the efficiency of testing.

Given limited testing time, improving testing efficiency means more test cases, leading to relatively high-quality apps.
Towards that target, there have been some works~\cite{hu2014efficiently,song2017ehbdroid,wang2021infrastructure} leveraging advanced infrastructure support to fetch GUI hierarchy and execute events efficiently.
Besides the infrastructure efficiency, the impact of GUI rendering has been overlooked.
GUI rendering is the act of generating a frame from the app and displaying it on the screen~\cite{web:render} including transiting from the last page, loading resources from internet, drawing UI objects (button) into pixels following the order of view hierarchy, etc, as seen in Fig.~\ref{fig:timeline}.
That process may be long depending on the app code quality, device performance and internet bandwidth.
Typically, automated testing tools configure a fixed amount of time (throttle) between events, in order to wait for the GUI to be fully rendered.
Setting an optimal throttle can help reduce the waiting time of automated testing tools, resulting in higher testing efficiency.

However, the fixed throttle may not work for different testing tools on different devices, and even different pages in the same app due to the screen complexity difference of each GUI.
Although long throttle can bring fully rendered GUIs, it may slow down the whole testing process for some idle waiting (e.g., Fig.~\ref{fig:timeline}E). 
If the throttle is too short, many GUIs may just be partially rendered which negatively affect the testing effectiveness~\cite{wang2017skyfire,choi2018detreduce,choi2013guided,negara2019practical} due to two major reasons.
First, there are many GUI testing tools highly dependent on visual information of GUI, including usability bug detection~\cite{liu2020owl}, robot testing~\cite{ran2022automated}, reinforcement-learning-based app exploration~\cite{adamo2018reinforcement}, test case migration across platforms~\cite{talebipour2021ui} which require fully rendered GUI as the input.
% Second, bugs triggered from partially rendered GUIs may not be encountered by real-world users \revise{(e.g., Fig.~\ref{fig:timeline}B and \ref{fig:timeline}C)}, and such bugs may hinder the following app exploration, resulting in the missing of other bugs.
% \chen{As discussed in our last meeting, please do not mention such reasons. Tell that since XML does not aligned well with GUI rendering, the action sends to partially-rendered GUI may not be executed as expected...}
% \chen{Revised, please double check. BTW, cannot see clearly the text in Fig.~\ref{fig:timeline}, may use PDF?}
Second, the run-time view hierarchy may be out of sync with the rendering GUI, and the action based on the view hierarchy may not be executed as expected, resulting in low coverage (e.g., tapping ``Screws'' image by coordinates from the view hierarchy file will be missed in rendering GUI at Fig.~\ref{fig:timeline}C).
%as there can be invisible objects on the screen, misaligned objects that only partially covering the rendered objects, or objects that are grayed out and cannot be executed (e.g., Fig.~\ref{fig:timeline}C).
%Those issues may hinder the following app exploration, resulting in the missing of bugs.
%First, some events triggered on transiting state (200ms at Fig.~\ref{fig:timeline}) are prone to invalid app crashes (would not be triggered when the GUI is fully rendered), resulting in redundant test sequences.
%Second, most events fail to trigger on non-interactive GUIs (e.g., sending events to a loading state in 400ms at Fig.~\ref{fig:timeline}), which restricts their capability of discovering deep app functionalities.
%On the other hand, if the throttle is too slow, the automated tool may stagnate on the GUI (800ms at Fig.~\ref{fig:timeline}), which reduces the efficiency of testing.
Therefore, an adaptive throttle (e.g., 600ms in Fig.~\ref{fig:timeline}) is needed to strike a balance between effectiveness and efficiency.
%Finding an adaptive throttle for different testing tools in different platforms is not a trivial task.

To further understand the throttling issues in automated testing tools, we first carry out a pilot study on 3 widely-used GUI testing tools for 32 apps to observe the GUI rendering.
Results show that a fixed short throttle setting (e.g., 200ms) causes 24\% of events on average happening on partially rendered states. %which can seriously impact the testing effectiveness.
The partially rendered states mainly include \textit{transiting state}, \textit{explicit loading state}, and \textit{implicit loading state}.
Although extending the throttle interval can help address the issues with a partially rendered state, an excessive long throttle (e.g., 1000ms) reduces 52.8\% testing events of automated exploration, which can seriously impact the testing efficiency.
These findings motivate this work in finding an adaptive throttle during GUI testing, and the difference between fully rendered and partially rendered lays the foundation for our approach.

%While the app under testing is mostly idle, the tool has to wait until the GUI finishes rendering before moving to the next event.
% One common practice to adaptive throttling is fetching the view hierarchy of the GUI, such as Appium\cite{web:appium}, Airtest~\cite{web:airtest}, etc.
% However, subsequent researches~\cite{li2020mapping,li2022learning,liu2018learning} have found that the fetched GUI views may out of sync, leading to tests on misaligned or invalid objects.
We propose a lightweight approach \tool, to automatically throttle the events adaptive based on GUI rendering inference.
Specifically, we formulate the throttle time prediction problem as a run-time classification task by discriminating between fully rendered and partially rendered GUIs.
We adopt a deep learning method to model the visual information from the GUI screenshot for inferring the GUI state.
First, we leverage image processing techniques to extract frames from GUI transiting screencasts to construct a large-scale binary GUI dataset, including 66,233 fully rendered and 45,623 partially rendered GUIs. 
Then, we adopt a small but efficient Convolutional Neural Network (CNN) based approach to discriminate the GUI rendering state.
To deploy our approach in testing tools to synchronize GUI rendering inference and schedule testing events, so as to send events until the GUI is fully rendered, we implement a socket-based framework to stream the real-time GUI screenshots and GUI rendering inference.
Note that one strength of our approach is that it is purely image-based, which can be easy to deploy in real-world practice.

To evaluate the accuracy of our \tool, we carry out a large-scale experiment on 20,125 GUI screenshots from 1,877 Android apps.
Compared with 11 state-of-the-art baselines, our \tool can achieve more than 99.8\% accuracy in predicting GUI rendering state.
We also conduct an experiment to demonstrate that our approach can speed up automated testing without sacrificing testing effectiveness by replaying 18 existing crash bugs from 12 defective Android apps.
%Results show that our \tool can detect all of the bugs in less run-time, compared to the benchmark and throttling methods.
In addition, the efficiency of current GUI testing tools is further boosted by novel approach design and implementation.
%To demonstrate the advantages of our \tool in real-world practice, we evaluate the usefulness of our approach by integrating into the automated testing tool.
Given the same run-time for the testing tool with and without \tool, the \tool-enhanced tool can achieve 6.96\% higher activity coverage and execute 13.24\% more events, than the vanilla tool.

% As the ultimate goal of automated testing is to detect bugs, we conduct an experiment on 12 defective Android apps.

The contributions of this paper are as follows:
\begin{itemize}
    \item To the best of our knowledge, this is the first study to automatically infer GUI rendering status for accelerating GUI testing. We propose a lightweight computer-vision based approach, \tool\footnote{\url{https://github.com/sidongfeng/AdaT}} to adaptively adjust the throttle between events.
    \item A motivational empirical study to understand GUI rendering process in Android apps to motivate this study and lays the foundation for our research. %of automated testing tools to identify throttling issues with balancing testing effectiveness and efficiency.
    \item Comprehensive experiments including the performance of \tool and its integration with the automated testing tool to demonstrate the accuracy, efficiency, effectiveness, and usefulness of our approach.
\end{itemize}
\section{Motivational Study}
\label{sec:motivation}
To better understand the issues of automated testing tools with throttling, we carried out a pilot study to examine the prevalence of these issues, so as to facilitate the development of our tool to enhance the existing Android testing tools.

% Existing works~\cite{web:monkey, gu2019practical,li2017droidbot} have been focusing on designing sophisticated GUI exploration algorithms to achieve better testing effectiveness.
% Although testing effectiveness can be improved with sophisticated algorithms, one often overlooked aspect is the efficiency of testing.
% Therefore, we carried out an empirical study to understand the sources of efficiency issues from automated testing, so as to facilitate the development of our tool to enhance the existing Android testing tools.
% To guide enhancements to Android testing tools, we conduct a motivating study to understand the extent and sources of inefficiency from automated testing.

\subsection{Experiment Setup}
We collected 32 Android apps as our experimental dataset, which were used in previous studies~\cite{feng2022gifdroid,feng2022gifdroid2,bernal2020translating}.
They are all top-rated on Google Play, covering different app categories such as news, tools, medical, etc.
Details of these apps are shown in our online appendix.
These apps do not require logging in, given that logins can be flaky which may affect the experimental measurement.
% To drive testing on these apps, we employed three widely-used Android testing tools, including Monkey, Ape, and Humanoid.
Each app was run for 3 minutes without interruption.
Note that we captured a GUI screenshot before triggering each event to visualize what GUI the event executed on.

\begin{comment}  
\sidong{may remove this paragraph for more space.}
All of our experiments were conducted on the official Android x86-64 emulators running Android 6.0 on a server. 
Each emulator was allocated with 4 dedicated CPU cores, 2GiB of RAM, and 2GiB of internal storage space.
The emulators were stored on a RAM disk and backed by discrete graphics cards for minimal mutual influences caused by disk I/O bottlenecks and CPU-intensive graphical rendering.
\end{comment}

\subsection{Categorizing GUI rendering state}
\label{sec:motivate_1}
To understand the GUI states in testing, we conducted a small pilot study on GUI screenshots collected by the commonly-used automated GUI testing tool Droidbot~\cite{li2017droidbot}, executing with throttle 200ms interval, which is a common setting in automated testing~\cite{patel2018effectiveness}.
% Note that we selected Monkey in our pilot study as it is the most widely-used app GUI testing tool in real-world practice.
% \revise{Note that we selected Droidbot as it achieves the highest testing coverage and more GUI states in previous study~\cite{wang2018empirical}.}
During the manual examination process, we noticed that there are different types of GUI rendering states, a categorization of these states would help clarify the issues in tools.
We recruited two students as annotators by the university’s internal slack channel and they were compensated with \$12 per hour. According to the pre-study background survey, they have labeled one UI/UX-related dataset (e.g., GUI element bounding box). 
To ensure accurate annotations, the process started with initial training.
First, we asked them to read a document~\cite{web:render} that outlines the GUI rendering process. 
Second, we provided an example set of annotated GUIs where the rendering states have been labeled by authors. 
This enforces a deeper understanding of the GUI rendering states. 
Third, we asked them to pass an assessment test, which includes a set of test GUIs.
Finally, we asked them to manually check 1,500 random GUIs and classified them into four categories following the Card Sorting~\cite{spencer2009card} method:

\textbf{Fully Rendered State.}
A fully rendered state represents a GUI rendered completely with all resources loaded and displayed.
%a complete transition to the GUI with all resources loaded.

\begin{figure}
	\centering
	\subfigure[Transiting state]{
		\includegraphics[width = 0.29\linewidth]{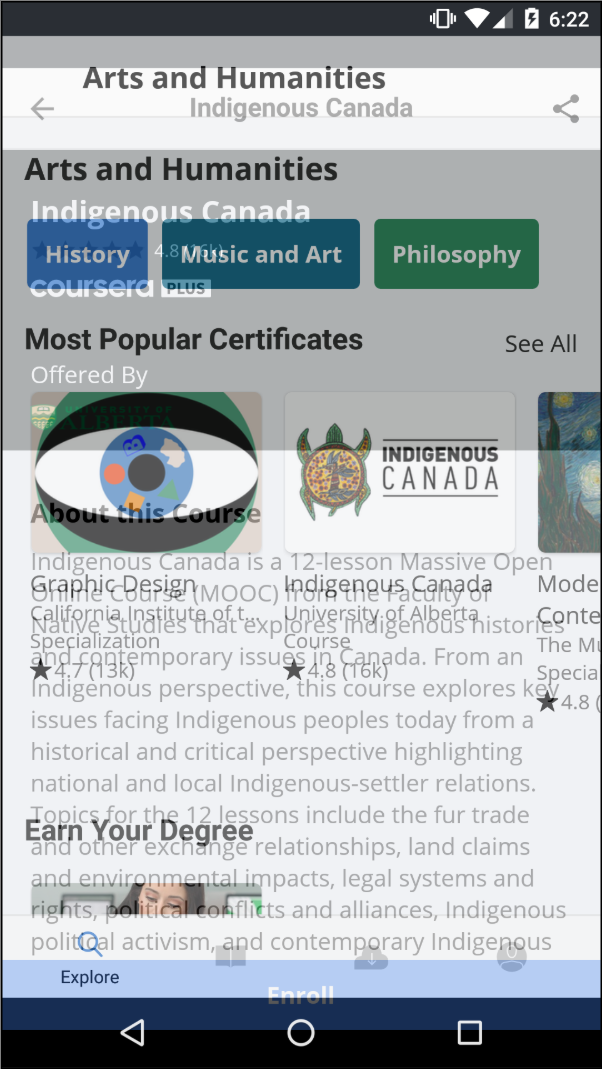}
		\label{fig:partial_1}}
	\hfill
	\subfigure[Explicit loading]{
		\includegraphics[width = 0.29\linewidth]{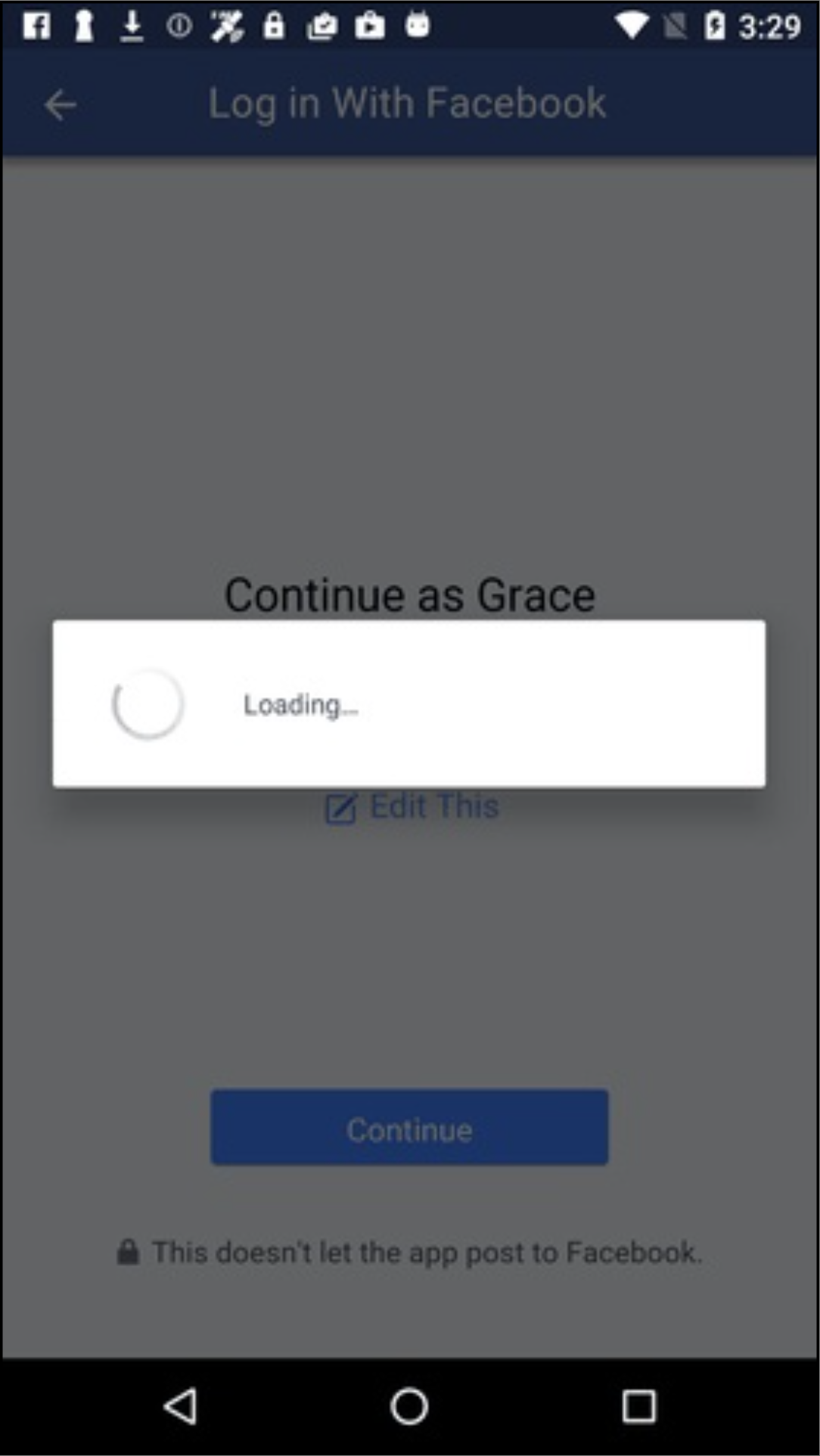}
		\label{fig:partial_2}}	
	\hfill
	\subfigure[Implicit loading]{
		\includegraphics[width = 0.29\linewidth]{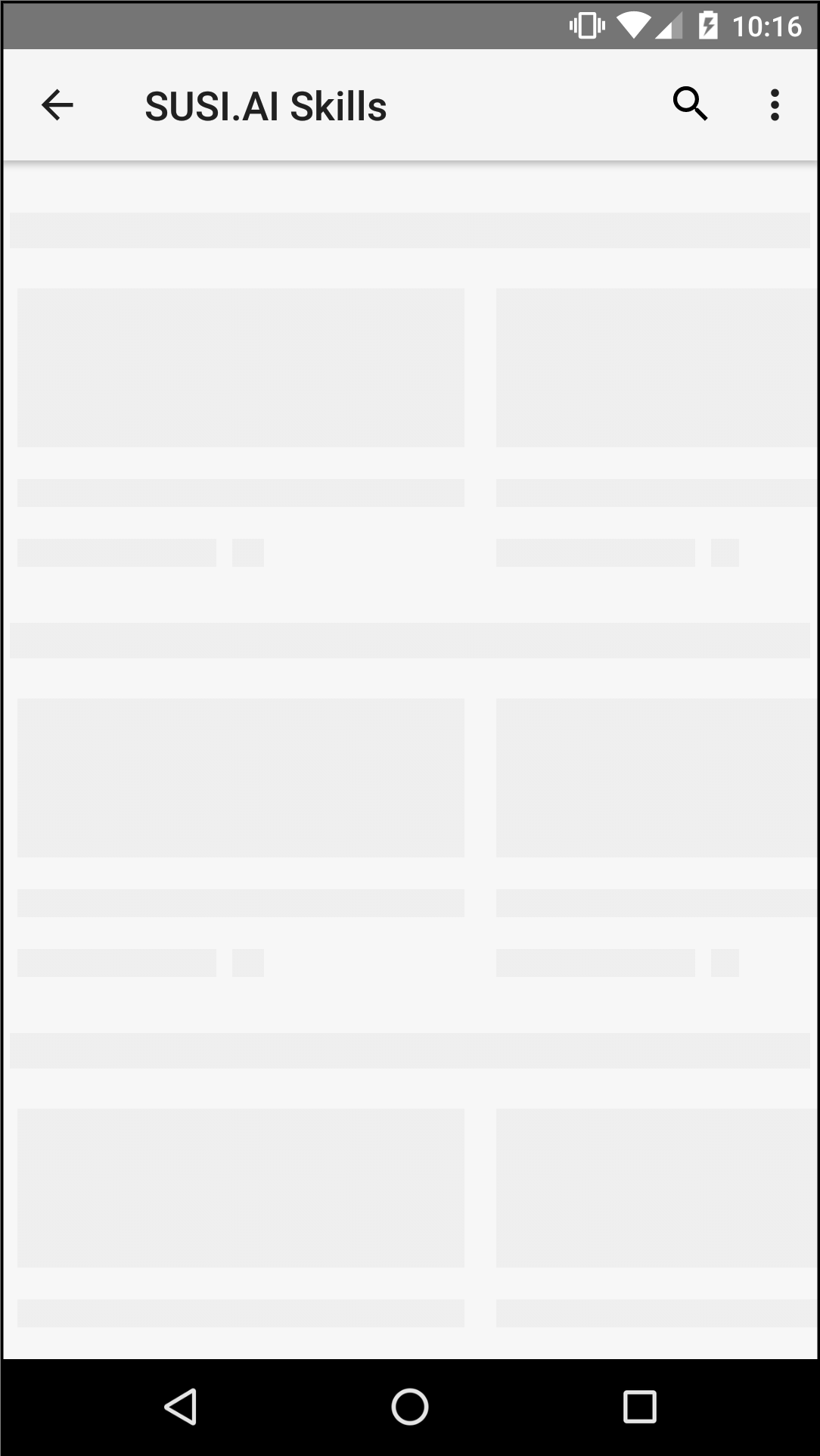}
		\label{fig:partial_3}}	
	\caption{Examples of partially rendered state.}
	\label{fig:partial}
\end{figure}

\textbf{Transiting State.}
As shown in Fig.~\ref{fig:partial_1}, one state is transiting to the next state.
As the transition between states takes longer than the throttle interval, two GUIs are overlapped with each other.
There are mainly two reasons for capturing transiting state.
First, the throttle setting is too short to get GUI fully rendered.
Second, there may be issues with the app development (e.g., too many animations, defects in the hardware acceleration), resulting in an unexpected long rendering process.
\begin{comment}
It might be caused by developers may employ some attractive transition effect between GUIs (similar to widget effect), which takes slightly longer to transit. 
It might be also due to defects in the hardware, or the exclusion of hardware acceleration during transition.
So when the next event is sent, the GUI of the previous state has not transited completely.
As a result, triggering events on misaligned objects can easily crash the app.
\end{comment}

\textbf{Explicit Loading State.}
As shown in Fig.~\ref{fig:partial_2}, it shows an explicit loading state, depicting a loading bar in the GUI, such as a spinning wheel, linear progressing bar, textual hint, etc.
It explicitly indicates the process or rendering is in progress and is often used for secure data transformations, such as logging accounts, transferring money, uploading a file, etc.
During the explicitly loading state, the GUI is non-interactive.

\textbf{Implicit Loading State.}
As shown in Fig.~\ref{fig:partial_3}, some resources are not showing due to network latency or resource defects.
Note that, for the explicit loading state, there is always a loading bar; while for the implicit loading state, the loading resources need to be recognized accordingly. 
For example, in Fig.~\ref{fig:partial_3}, the loading resources appear as some gray layouts.
% In addition to the ineffectiveness issues with explicit loading, the events that execute on the implicit loading resources may throw an unhandled exception, causing the app to crash.
% \revise{Bug triggered from implicit loading state may not be encountered by real-world users.}

% For example, in Firefox browser, tab switching could take up to ten seconds in certain scenarios
% Paper: Characterizing and Detecting Performance Bugs for Smartphone Applications

\noindent\fbox{
	\parbox{0.95\linewidth}{
		\textbf{Summary}: By conducting a pilot study on GUIs collected by Droidbot, we categorize four types of GUI rendering states that lie into fully rendered states, and partially rendered states (e.g., transiting state, explicit loading state, and implicit loading state)
	}
}

\subsection{Are partially rendered states common in testing tools?}
\label{sec:motivate_2}
To investigate whether testing on partially rendered states is ubiquitous in existing tools, we audited the GUI screenshots captured from three commonly-used testing tools, including Droidbot~\cite{li2017droidbot}, Monkey~\cite{web:monkey}, and Ape~\cite{gu2019practical}, with 200ms throttle interval.
In total, we obtained 875, 2,830, and 1,646 GUI screenshots from Droidbot, Monkey, and Ape, respectively.
According to our GUI categories observed in Section~\ref{sec:motivate_1}, the two annotators first annotated the GUI screenshots independently without any discussion, and then met and discussed the discrepancies until consensus was reached.

\begin{figure}
	\centering
	\includegraphics[width=0.95\linewidth]{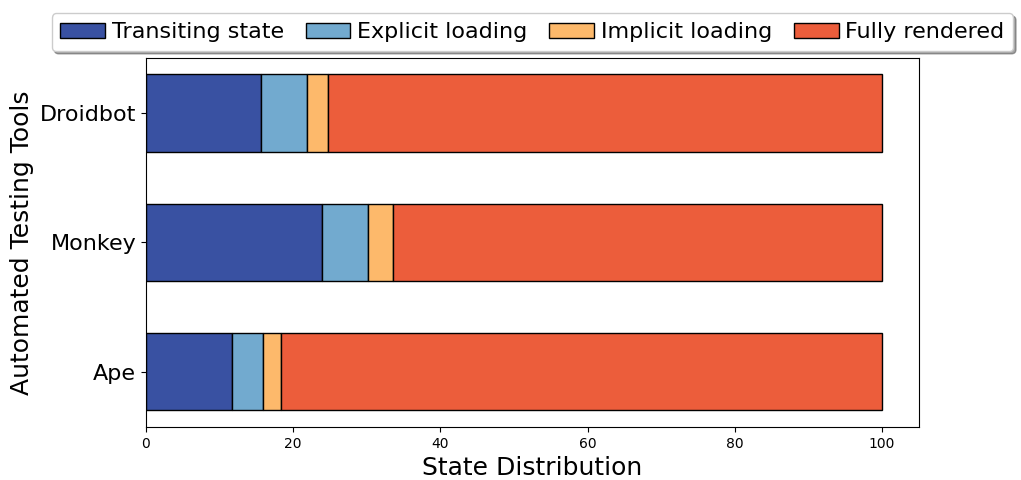}
	\caption{Distribution of rendering states captured by Droidbot, Monkey, and Ape.}
	\label{fig:occupation}
\end{figure}

Fig.~\ref{fig:occupation} depicts the results, showing percentages of GUI state categories in testing tools.
We can see that all of the testing tools have the issues of testing events on partially rendered states, i.e., 23\%, 32\%, 17\% exist in Droidbot, Monkey, and Ape, respectively.
These partially rendered states will potentially reduce the testing effectiveness.
As we can see that state transition is a major partially rendered problem arises in automated testing tools, e.g., 15\%, 23\%, 11\%, in Droidbot, Monkey, and Ape.
This indicates that some GUI transitions need more time (longer than 200ms) to finish the transition.
The consecutive actions by the testing tool to the incomplete rendered GUI may not trigger the expected event, resulting in a decrease of testing coverage.

\noindent\fbox{
	\parbox{0.95\linewidth}{
		\textbf{Summary}: By analyzing three commonly-used testing tools, we find that they all encounter the issue with partially rendered states, which may negatively influence the effectiveness when testing.
	}
}

\subsection{How to avoid partially rendered states?}
To address the issue of partially rendered GUIs, the simplest way is to set a longer throttle interval, extending the inter-event time for transiting or loading.
Therefore, we investigated how throttle affects the testing tool by running Droidbot with 5 different throttle intervals, including 200ms, 400ms, 600ms, 800ms, and 1000ms.
% Note that we selected Ape because it achieved the best performance in Section~\ref{sec:motivate_2}.
We further annotated the GUI screenshots following the procedure in Section~\ref{sec:motivate_2}.

\begin{figure}
	\centering
	\includegraphics[width=0.95\linewidth]{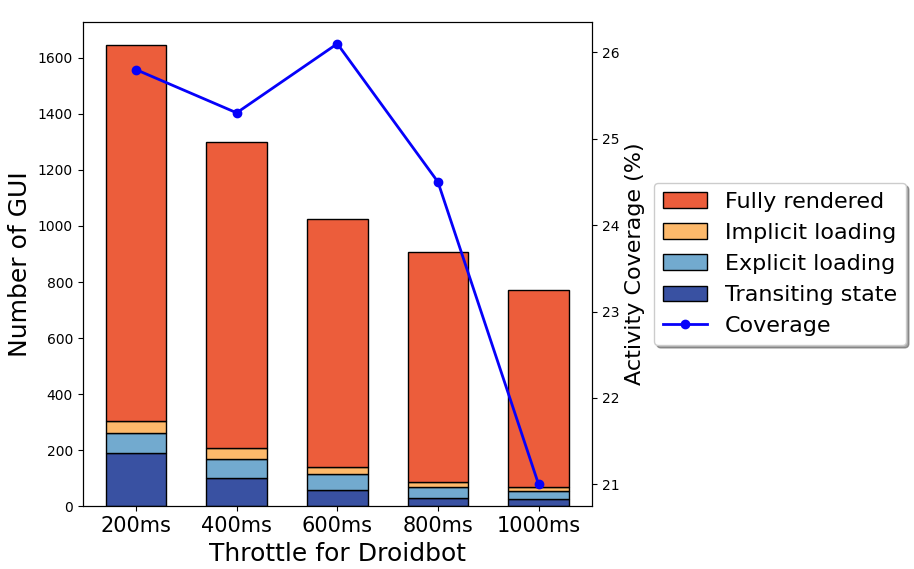}
	\caption{Number of GUIs and activity coverage in different throttle settings of Droidbot.}
	\label{fig:throttle}
\end{figure}

Fig.~\ref{fig:throttle} depicts the results, showing the number of GUIs and the activity coverage.
We can observe that by extending the throttle intervals, the issue with partially rendered states is mitigated, i.e., 17\%, 15\%, 14\%, 9\%, 8\% incomplete rendering for the throttles from 200ms to 1000ms, respectively.
Specifically, the issue of transiting states substantially decreases.
This indicates the GUIs can ease of transiting and loading between events with longer intervals.
However, utilizing longer throttle intervals decreases the number of GUIs, e.g., 1,646, 1,299, 1,023, 907, 776 GUIs for throttles from 200ms to 1000ms, indicating fewer events executed at the run-time.
Consequently, the activity coverage constantly drops, except for the 600ms throttle.
This is because when the testing is over-stressed (e.g., 200ms, 400ms), the tool may encounter the issues of the partially rendered state, leading to ineffective testing; 
when the test throttling is appropriate (e.g., 600ms), the tool sends events to fully rendered GUIs, exploring more activities;
when the testing is less-stressed (e.g., 800ms, 1000ms), the tool may stagnate testing, leading to inefficient testing.
Therefore, a suitable throttle should strike a balance between effectiveness and efficiency.

\noindent\fbox{
	\parbox{0.95\linewidth}{
		\textbf{Summary}: By analyzing five different throttle intervals, we find that extending throttle can help address the issue with partially rendered states.
		However, an excessive long throttle can reduce the efficiency of automated exploration.
	}
}

\subsection{Why makes throttle adaptive?}
These findings confirm the importance of throttle setting to automated testing, and motivate us to design an approach for balancing effectiveness and efficiency.
While the app under testing is mostly idle, the tool has to wait until the GUI finishes rendering before moving to the next event.
Taken in this sense, it is worthwhile developing a new effective and efficient method to dynamically adjust the throttle during testing.
The underlying issue is to infer GUI rendering states, discriminating partially rendered GUI and fully rendered GUI.
Inspired by the fact that these GUIs can be easily classified by human eyes, we propose to identify the GUI rendering states with visual cognitive techniques.
As the GUI screenshots are easy to capture for all automated testing tools, our image-based approach is more general and easier to deploy.
% , compared with the program-analysis based method.

\section{Approach}

\begin{figure}
	\centering
	\includegraphics[width=0.95\linewidth]{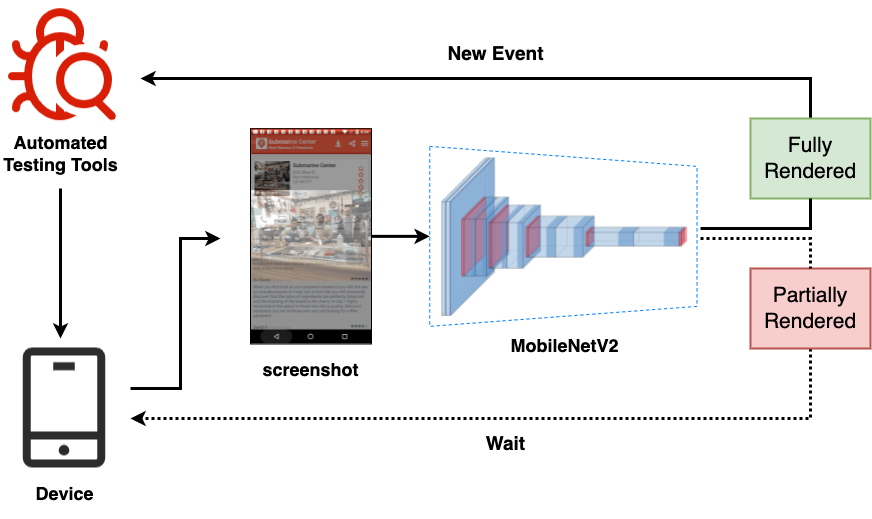}
	\caption{Overview of our approach.}
	\label{fig:overview}
\end{figure}

This paper proposes a simple but effective approach \tool to adaptively adjust the throttle base on GUI screenshots.
Given that automated testing tools test on the device, we synchronously stream the GUI screenshot capturing, and detect its current rendering state.
Based on the GUI rendering inference, we schedule the testing events, which will be sent if the GUI is fully rendered, otherwise, wait explicitly for rendering.
The overview of our \tool is shown in Fig.~\ref{fig:overview}.

% Learning rate schedules seek to adjust the learning rate during training by reducing the learning rate according to a pre-defined schedule.
The fundamental of \tool is to adopt a lightweight CNN-based model to classify the GUI rendering state, which is divided into three main phases:
(i) the \textit{Data Preparation} phase, which automatically collects a large-scale dataset of partially rendered GUIs and fully rendered GUIs,
(ii) the \textit{GUI Rendering State Classification} phase that proposes a CNN-based model to discriminate the current GUI rendering state,
and (iii) the \textit{Model Deployment} phase that proposes an efficient deployment of our model in testing tools.

% jank
% https://developer.android.com/topic/performance/vitals/render

\subsection{Data Preparation}
\label{sec:phase1}
The foundation of understanding GUI rendering state and training deep learning model is big data, whereas manual labeling is prohibitively expensive.
The goal of this phase is to automatically collect partially rendered GUIs and fully rendered GUIs by leveraging GUI transiting screencasts, as shown in Fig.~\ref{fig:dataset}.

\subsubsection{GUI Transiting Screencasts}
\label{sec:phase1_1}
We use the open-sourced Rico dataset~\cite{deka2017rico}, which contains 44,418 transiting screencasts from more than 9.7k different Android applications in 27 different app categories.
The duration of screencasts spans from 0.5 to 50 seconds.
Each screencast contains one or multiple user actions (e.g., tap, scroll) and no-action periods.
We discard the transiting periods of scroll action in our dataset.
This is because the GUI state can be ambiguous on the development mechanism.
For example, scrolling on a lazy-loading GUI may collect partially rendered GUIs; scrolling a pre-loaded GUI may collect fully rendered GUIs.
Finally, we obtain 36,038 transiting screencasts.

\subsubsection{Transiting Frame Identification}
% \chen{animation ??I think ``transiting/transition'' is better than animation which is more about fancy dynamic visual effect. If agreed, please update the words in the paper for consistency.}
A GUI transition is comprised of frames of partially rendered and fully rendered.
To identify the frame state in the transiting screencast, we adopt an image processing technique to build a perceptual similarity score for consecutive frame comparison based on Y-Difference (or Y-Diff).
YUV is a color space usually used in video encoding, enabling transmission errors or compression artifacts to be more efficiently masked by the human perception than using a RGB-representation~\cite{chen2009compression,sudhir2011efficient}.
Y-Diff is the difference in Y (luminance) values of two images in the YUV color space, used as a major input for the human perception of motion~\cite{livingstone2002vision}.

Consider a transiting screencast $\big\{ f_{0}, f_{1}, .., f_{N-1}, f_{N} \big\}$ , where $f_{N}$ is the current frame and $f_{N-1}$ is the previous frame.
To calculate the Y-Diff of the current frame $f_{N}$ with the previous $f_{N-1}$, we first obtain the luminance mask $Y_{N-1}, Y_{N}$ by splitting the YUV color space converted by the RGB color space.
Then, we apply the perceptual comparison metric, Structural Similarity Index (SSIM)~\cite{wang2004image}, to produce a per-pixel similarity value related to the local difference in the average value, the variance, and the correlation of luminances.
A SSIM score is a number between 0 and 1, and a higher value indicates a strong level of similarity.

\begin{figure}
	\centering
	\includegraphics[width=0.95\linewidth]{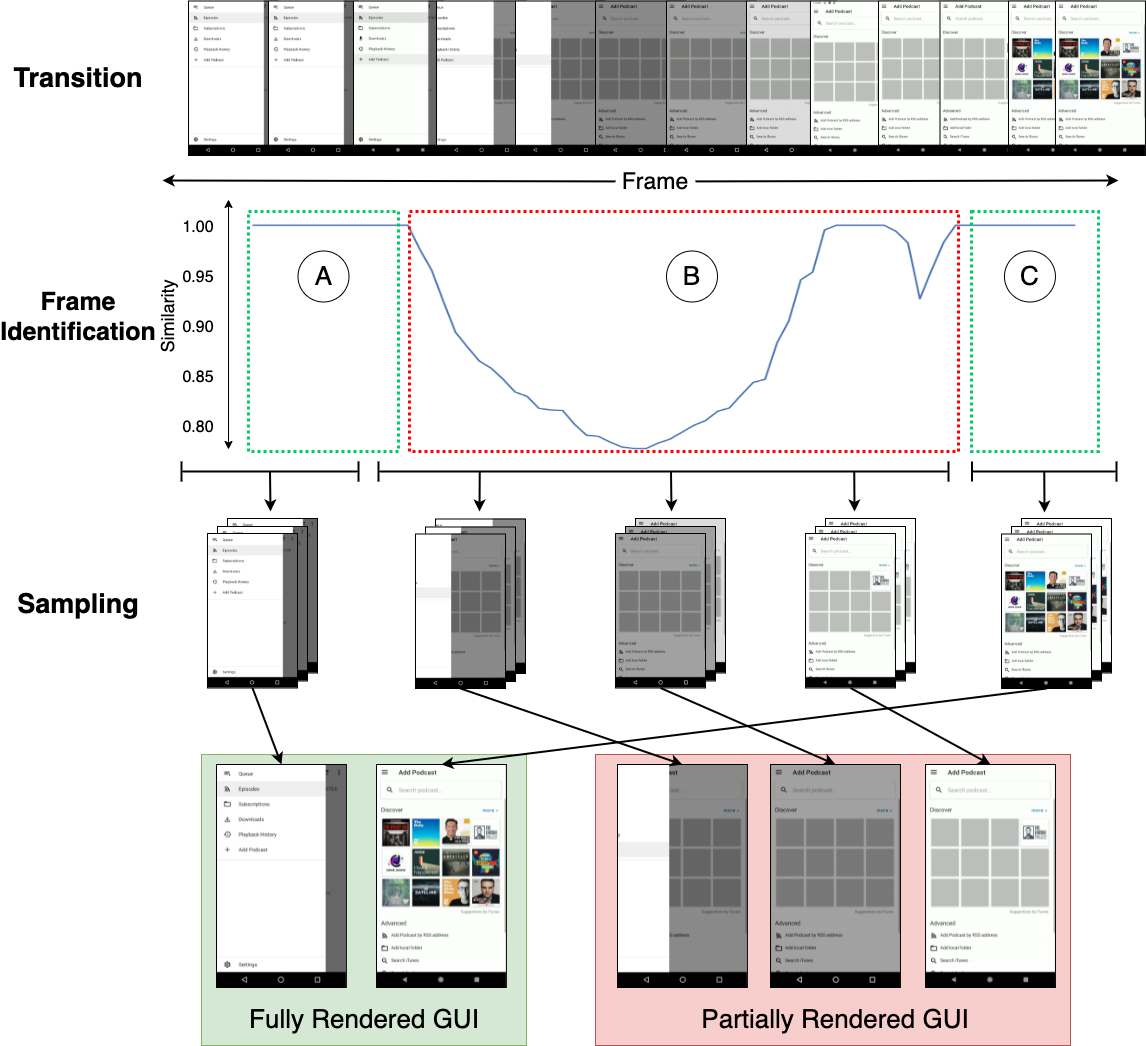}
	\caption{Pipeline for automated data collection.}
	\label{fig:dataset}
\end{figure}

To identify whether one frame is fully or partially rendered, we look into the similarity scores of consecutive frames in the transiting screencast as shown in Fig.~\ref{fig:dataset}.
The first step is to group frames belonging to the same atomic state according to a tailored pattern analysis.
This procedure is necessary because discrete states performed on the screen will persist across several frames, and thus, need to be grouped and segmented accordingly.
We find that a fully rendered GUI is in a steady state where the consecutive frames are the same or very similar for a relatively long duration, for example, Fig.~\ref{fig:dataset} (A) and (C).
In contrast, a partially rendered GUI shows a great difference on the consecutive frames, revealing an instantaneous transition from one screen to another.
For example, as shown in Fig.~\ref{fig:dataset} (B), when the user clicks a button, the current GUI starts to fade out, in which the similarity score starts to drop drastically. Afterwards, the next GUI starts to fade in and the similarity score rises.
According to our observation, one common case in partially rendered GUI is that the similarity score becomes steady for a small period of time between two drastically droppings as shown in Fig.~\ref{fig:dataset} (B). 
The occurrence of this short steady duration is because of the resource loading in GUI, aligning with our observation of implicit loading state in Section~\ref{sec:motivate_1}.
We have empirically set 0.99\footnote{We set up that value by a small-scale pilot study} as the threshold to decide whether two frames are similar, and 5 frames as the threshold to indicate a steady state, in order to differentiate fully rendered and partially rendered GUIs.

\subsubsection{GUI State Sampling}
% \chen{Why K=3? This part is a bit complicated, can we make it simple, just consecutively sampling part of data for data diversity and balance.}
For each GUI state group, we observe that although the frames are densely recorded in the screencasts, the rendering changes relatively slowly.
To prevent bias on redundant data, we propose an approach to sample the GUI frames from GUI groups.
As the GUI frames in the fully rendered group are similar or identical, we randomly select one frame.
To sample frames from partially rendered groups, we adopt a paradigm using uniform sampling~\cite{shi2013sampling} to ensure the diversity in partially rendered GUIs and to prevent the bias of imbalance sampling between fully rendered and partially rendered GUIs.
After automated identifying and sampling, we obtain a dataset with 66,233 fully rendered GUIs and 45,623 partially rendered GUIs.

\begin{comment}
To sample frames from partially rendered groups, we adopt a paradigm using K-means clustering~\cite{likas2003global}. 
This strategy is based on sparsely and distinctly sampling.
We first extract the features from each GUI frame in the partially rendered group using grayscale pixel values. 
Then, we initialize $K$ centroids and keep assigning features to a cluster such that the sum of the squared distance between the features and the cluster’s centroid until minimum.
We select the $K$ centroids as our partially rendered samples.
Concerning the property of spareness and preventing the bias of imbalance sampling from fully rendered and partially rendered~\cite{chawla2009data}, we empirically set $K$ to 3.
\end{comment}

\subsection{GUI Rendering State Classification}
\label{sec:phase2}
In this phase, we identify whether the GUI is fully rendered which allows testing tools to execute the next event.
%or whether the GUI is partially rendered which waits until the rendering is complete.
%Since the ultimate goal of our tool is to improve the efficiency of the testing tools, the classification of GUI state aims to be lightweight.
%Advanced deep learning models leveraging Convolutional Neural Networks (CNNs) for precise and efficient image recognition~\cite{krizhevsky2012imagenet,lecun1998gradient} have reached human levels of accuracy for image classification tasks.
To differentiate between fully rendered and partially rendered GUIs, we adopt an implementation of MobileNetV2~\cite{sandler2018mobilenetv2}, which distills the best practices in convolutional network design into a simple architecture that can serve as competitive performance but keep low parameters and mathematical operations to reduce computational cost and memory overhead.
In addition to the simulator, the model can even be deployed on mobile devices for efficient testing.
This advanced network design speeds up image classification, which is the ultimate goal of this work to efficiently discriminate the GUI rendering states.
% \chen{Neet to highlight 3 important attributes: 1)good performance 2)fast 3)low memory overhead, can run on devices for GUI testing.}
% \sidong{I suggest not to add device testing, which may confuse reviewers.}

Specifically, we adopt a more advanced depthwise separable convolution, combining one $3*3$ convolution layer and two $1*1$ convolution layers to capture essential information from images.
We first use a $1*1$ pointwise convolution layer to expand the number of channels in the input feature map.
Then, we use a $3*3$ depthwise convolution layer to filter the input feature map and a $1*1$ convolution layer to reduce the number of channels of the feature map.
In order to improve the performance and stability between layers, we borrow the idea of residual connection in ResNet~\cite{he2016deep} to help with the flow of gradients.
After the convolution layer, we add Batch Normalization (BN)~\cite{ioffe2015batch} to standardize the feature map.
Finally, the activation function, Rectified Linear Unit (ReLU), is added to increase the nonlinear properties of the classifier function and of the overall network without affecting the features.

For detailed implementation, we adopt the stride of 2 in the depthwise convolution layer to downsample the feature map.
We use ReLU6 defined as $y=min(max(0,x),6)$, for the first two activation layers because of its robustness in low-precision computation~\cite{howard2017mobilenets}.
A linear transformation (also known as Linear Bottleneck Layer)~\cite{sandler2018mobilenetv2} is applied to the last activation layer to prevent ReLU from destroying features.
The momentum in the BN layer is set as 0.1.
To make our training more stable, we adopt Adam as optimizer~\cite{kingma2014adam}, and binary CrossEntropyLoss as the loss function~\cite{murphy2012machine}.
Moreover, to optimize the training model, we apply an adaptive learning scheduler, with an initial rate of 0.01 and decay to half after 10 iterations.
For data preprocessing, we resize the GUI screenshots to $768*448$. 
We implement our model based on the PyTorch framework~\cite{paszke2019pytorch}.
Note that the hyper-parameter settings are determined empirically by a small-scale experiment.
%were selected after an initial set of experiments.

\subsection{Model Deployment}
\label{sec:phase3}
\begin{figure}
	\centering
	\includegraphics[width=0.95\linewidth]{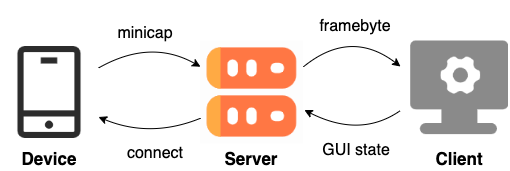}
	\caption{Overview of model deployment.}
	\label{fig:implementation}
\end{figure}

To make the model efficiently provide feedback on the GUI rendering state to the automated testing tool, synchronization of the GUI and the testing tool is needed.
However, capturing and transmitting GUI screenshots can be time-consuming.
Therefore, we develop a socket-based smartphone test farm using OpenSTF~\cite{web:openstf} to stream the real-time GUI screenshot.
It is a framework to facilitate the mobile testing process by accessing mobile devices remotely.
In detail, the framework consists of three components: \textit{Device Side}, \textit{Server Side}, and \textit{Client Side}. 
Each component leverages fast and safe microservices to communicate with each other, such as ZeroMQ~\cite{web:zeromq} and Protocol Buffer~\cite{web:protocol}.
The overview of the model deployment is shown in Fig.~\ref{fig:implementation}.

The goal of \textit{Device Side} is to monitor and send events to the device as a background process.
We utilize the mature and efficient binary method Minicap~\cite{web:minicap}, to capture screenshots on the device. 
In detail, the screenshots are stored as a binary format, where the first 4 bits represent the screenshot size $n$, and the next $n$ bits represent the screenshot buffer, for accelerating data transfer between \textit{Device Side} and \textit{Server Side}.
The \textit{Server Side} is to keep a device tracker (e.g., daemon) to manage whenever a device is connected or if the device gets disconnected.
The \textit{Client Side} leverages the WebSocket to keep receiving the screenshot buffer from the server.

Once the screenshot buffer is received, we decode it into a PyTorch tensor~\cite{paszke2019pytorch}.
This tensor is then fed into our trained GUI state classification model to infer the rendering state of the current GUI.
If it is fully rendered, we continue to test on the new event, otherwise, we explicitly wait for the next screenshot buffer.
To prevent excessive time budget due to the long duration of resource loading or wrong prediction of our model, we set up a maximum waiting threshold.
The waiting threshold is empirically set as 1000ms by a small pilot study.
We make the model and the source code used to set up \tool publicly available\footnote{\url{https://github.com/sidongfeng/AdaT}}.

\section{Evaluation}
The main quality of our study is the extent to whether our \tool can effectively and efficiently accelerate the automated testing process.
To achieve our study goals, we formulate the following three research questions:

\begin{itemize}
    \item \textbf{RQ1}: How accurate and efficient is our model in classifying GUI rendering state?
	\item \textbf{RQ2}: How effective and efficient is our approach in triggering bugs?
	\item \textbf{RQ3}: How useful is our approach when integrated in real-world automated testing tools?
\end{itemize}

For \textbf{RQ1}, we present some general performance of our model for GUI rendering inference and the comparison with state-of-the-art baselines.
For \textbf{RQ2}, we carry out experiments to check if our tool can speed up the automated GUI testing, without sacrificing the effectiveness of bug triggering.
% However, the randomness of the automated testing may affect the efficiency measurement, that is exploring different objects across different runs.
% To ensure the validity of evaluation, we set up a bug seed (\texttt{$--$seed}), which would generate the same sequence of events across different runs to trigger bugs.
For \textbf{RQ3}, we integrate \tool with DroidBot as an enhanced automated testing tool to measure the ability of our approach in real-world testing environments.

\subsection{RQ1: Performance of Model}
\textbf{Experimental Setup.}
To answer RQ1, we first evaluated the ability of our model MobileNetV2 (in Sec.~\ref{sec:phase2}) to accurately and efficiently differentiate between fully rendered GUIs and partially rendered GUIs.
To accomplish the evaluation, we followed the procedure to generate the dataset outlined in Section~\ref{sec:phase1}.
Regarding our training-testing data split, a simple random split cannot evaluate the model generalizability, as the GUIs in the same app may have very similar visual appearances.
To avoid this data leakage problem~\cite{kaufman2012leakage}, we split the screens in the dataset by app, completing with the 8:1:1 app split for the training, validation, and testing sets, respectively.
The resulting split has 79k GUIs in the training dataset, 10k GUIs in the validation dataset, and 10k GUIs in the testing dataset.
The model was trained in an NVIDIA GeForce RTX 2080Ti GPU (16G memory) with 20 epochs for about 3 hours.

\textbf{Metrics.}
Since we formulated our problem as an image classification task, we adopted three widely-used metrics i.e., precision, recall, F1-score, to evaluate the accuracy of the model inference.
Precision is the proportion of GUIs that are correctly predicted as fully rendered among all GUIs predicted as fully rendered.

\eq{$$ precision = \frac{\#GUIs \ correctly \ predicted \ as \ fully \ rendered}{\#All \ GUIs \ predicted \ as \ fully \ rendered} \\ $$}
Recall is the proportion of GUIs that are correctly predicted as fully rendered among all fully rendered GUIs.

\eq{$$ recall = \frac{\#GUIs \ correctly \ predicted \ as \ fully \ rendered}{\#All \ fully \ rendered \ GUIs} \\  $$}
F1-score (F-score or F-measure) is the harmonic mean of precision and recall, which combine
both of the two metrics above.

\eq{$$ F1-score = \frac{2 \times precision \times recall}{precision + recall} $$}
For all metrics, a higher value represents better performance.
Since the ultimate goal is to speed up testing process, we also measured the time for inference.
For the inference time, a lower time cost represents faster inference of the GUI rendering state.

% The higher the accuracy score, the better the model can discriminate the GUI rendering state.
% To prevent additional computation cost from the invocation of initializing GPU~\cite{web:warmup}, we warmed up the GPU by randomly running over 1000 samples.

\begin{table}
    \small
    \centering
	\caption{Performance comparison with baselines}
	\label{tab:rq1}
	\begin{tabular}{l|c|c|c|c}
	\hline
	\bf{Methods} & \bf{Precision} & \bf{Recall} & \bf{F1-score} & \bf{Time (ms)}\\ 
	\hline
	SIFT+SVM & 0.763 & 0.755 & 0.758 & 15.81\\
    SIFT+KNN & 0.624 & 0.645 & 0.634  & 1.94\\
	SIFT+RF & 0.676 & 0.663 & 0.669 & 1.71\\
	\hline
	SURF+SVM & 0.711 & 0.723 & 0.716 & 16.94\\
    SURF+KNN & 0.601 & 0.666 & 0.631 & 1.27\\
	SURF+RF & 0.650 & 0.675 & 0.662 & 1.29\\
	\hline
    ORB+SVM & 0.674 & 0.736 & 0.703 & 18.10\\
    ORB+KNN & 0.601 & 0.642 & 0.620 & 1.15\\
	ORB+RF & 0.635 & 0.657 & 0.645 & 1.46\\
	\hline
    CNN & 0.863 & 0.816 & 0.838 & 38.10\\
    % ResNet & 0.986 & 0.976 & 0.981 & 51.74\\
% 	\hline
	\bf{\tool} & \bf{0.999} & \bf{0.996} & \bf{0.998} & \bf{43.02} \\
	\hline
	\end{tabular}
\end{table}

\textbf{Baselines.}
We set up 10 baseline methods, including machine learning-based and deep learning-based, that are widely used in image classification tasks as the baselines to compare with our model.
The machine learning-based methods first extract visual features from the GUI screenshots, and then employ a machine learner for the classification.
The deep learning-based methods use a convolutional neural network to extract the visual features and then utilize fully connected perceptrons for classification.

In detail, we adopted three types of feature extraction methods used in machine learning, e.g., Scale invariant feature transform (SIFT)~\cite{lowe2004distinctive}, Speed up robot features (SURF)~\cite{bay2006surf}, and Oriented fast and rotated brief (ORB)~\cite{rublee2011orb}.
With these features, we applied three commonly-used machine learning classifiers, e.g., Support Vector Machine (SVM)~\cite{kotsiantis2007supervised}, K-Nearest Neighbor (KNN)~\cite{keller1985fuzzy}, and Random Forests (RF)~\cite{breiman2001random}, for classifying the GUI rendering state.
The combination of three types of image features and three classification learning algorithms generated a total of 9 baselines. 
We also experimented with off-the-shelf feature extraction methods used in deep learning, e.g., traditional CNN with 3 convolutional layers~\cite{o2015introduction}.
% We used two fully connected layers as a deep learning classifier, and set the number of neurons in each layer to 32, and 2, respectively.
We set the number of neurons in fully connected layers to 2, representing whether the GUI is in a fully-rendered or partially-rendered state.
We trained the baselines following the same procedure of our approach.

\textbf{Results.}
Table~\ref{tab:rq1} depicts the performance of our model MobileNetV2 in classifying the fully rendered GUIs and partially rendered GUIs.
The performance of our model is much better than that of other baselines, i.e., 30.9\%, 31.9\%, 31.6\% boost in recall, precision, and F1-score compared with the best machine learning baseline (SIFT+SVM).
We observe that the methods based on deep learning perform much better than machine learning due to the reason that the machine learning lacks of feature introspection, as the feature of GUI rendering state varies.
Compared with the deep learning baseline, our model further improves 13.6\%, 18\%, 16\% in recall, precision, and F1-score, respectively.
In addition, our model takes on average 43.02ms per GUI inference, representing the ability of our model to accurately and efficiently discriminate the GUI rendering state.

\begin{figure}
	\centering
	\subfigure[]{
		\includegraphics[width = 0.21\linewidth]{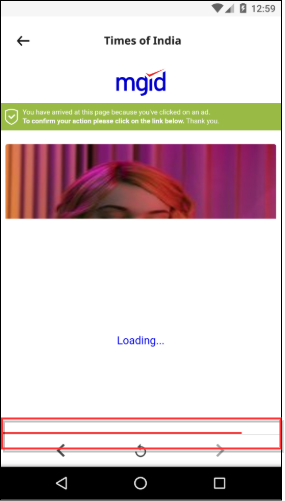}
		\label{fig:bad_1}}
	\hfill
	\subfigure[]{
		\includegraphics[width = 0.21\linewidth]{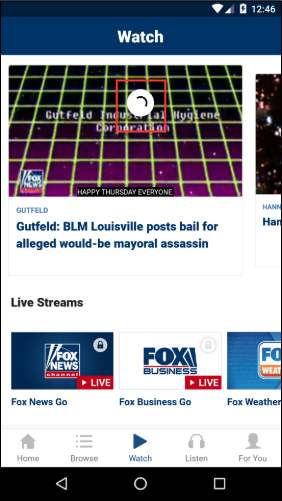}
		\label{fig:bad_2}}	
	\hfill
	\subfigure[]{
		\includegraphics[width = 0.21\linewidth]{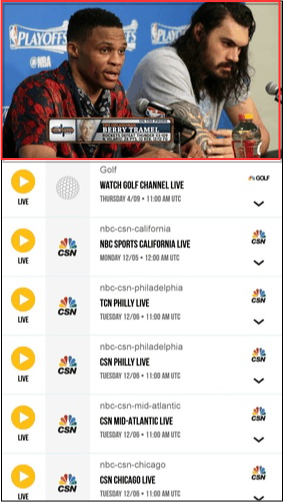}
		\label{fig:bad_3}}	
	\hfill
	\subfigure[]{
		\includegraphics[width = 0.21\linewidth]{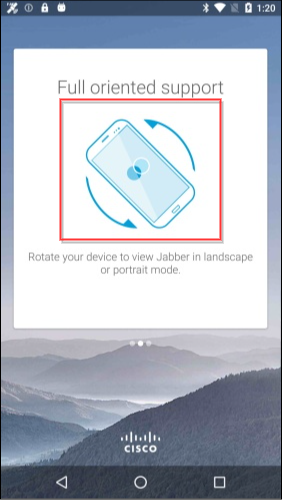}
		\label{fig:bad_4}}	
	\caption{ Examples of bad cases in GUI state prediction.}
	\label{fig:bad}
\end{figure}

% \chen{Try to make this shorter.}
Albeit the good performance of our model, we still make wrong predictions for some GUI screenshots. 
We manually check those wrong GUI cases and summarise two common causes.
First, within some GUIs, the representative features are too tiny and inconspicuous to be recognized even with human eyes, for example, the red linear progress bar in Fig.~\ref{fig:bad_1}, and the tiny circular progress bar embedded in the image in Fig.~\ref{fig:bad_2}.
Second, some negative data are not really negative data due to the different cognition of GUI state.
For example, since Fig~\ref{fig:bad_3} and \ref{fig:bad_4} are screenshots that contain dynamic assets such as videos and gifs, they are automatically annotated as partially rendered GUIs in temporal, while they seem to be fully rendered in static.
% Third, within some transitions, the resource loading is so slow that the partially rendered GUI may stay for a relatively long time, beyond our threshold setting. 
% So, that frame is wrongly annotated as a fully rendered GUI.

\begin{table*}
	\centering
	\scriptsize
	\caption{Performance comparison for our tool. ``T'' denotes the time to trigger the crash in seconds. ``R'' denotes crash reproducibility.
% 	\chen{For the last 4 approaches, all bugs can be successfully triggered? Then the success rate of our approach cannot be distinguished.}
	}
	\label{tab:rq2}
	\tabcolsep=0.12cm
    \begin{tabular}{l|r||c|c|c|c|c|c|c|c|c|c|c|c||c|c|c|c|c|c} 
    \hline
        \multirow{2}{*}{\bf{Crash Bug}} & \multirow{2}{*}{\bf{Step}} & \multicolumn{2}{c|}{\makecell{\bf{Throttle} \\ \bf{200ms}}} & \multicolumn{2}{c|}{\makecell{\bf{Throttle} \\ \bf{400ms}}} & \multicolumn{2}{c|}{\makecell{\bf{Throttle} \\ \bf{600ms}}} & \multicolumn{2}{c|}{\makecell{\bf{Throttle} \\ \bf{800ms}}} & \multicolumn{2}{c|}{\makecell{\bf{Throttle} \\ \bf{1000ms}}} & \multicolumn{2}{c||}{\bf{Themis}~\cite{su2021benchmarking}} & \multicolumn{2}{c|}{\makecell{\bf{Consecutive} \\ \bf{Frame}}} & \multicolumn{2}{c|}{\makecell{\bf{Asynchronous} \\ \bf{Streaming}}} &  \multicolumn{2}{c}{\cellcolor{lightgray} \bf{{\tool}}} \\
        \cline{3-20} &
        & T & R & T & R & T & R & T & R & T & R & T & R & T & R & T & R & \cellcolor{lightgray} T & \cellcolor{lightgray} R \\
    \hline
        \it{ActivityDiary\#118} & 10 & 12.77 & \cmark & 15.45 & \cmark & 17.31 & \cmark & 18.68 & \cmark & 20.54 & \cmark & 26.80 & \cmark & 21.74 & \cmark & 17.56 & \cmark & \cellcolor{lightgray} 16.68 & \cellcolor{lightgray} \cmark \\
        
        \it{ActivityDiary\#285} & 10 & 13.99 & \xmark & 15.79 & \cmark & 17.22 & \cmark & 18.97 & \cmark & 21.31 & \cmark & 34.49 & \cmark & 24.11 & \cmark & 18.66 & \cmark & \cellcolor{lightgray} 15.80 & \cellcolor{lightgray} \cmark \\
        
        \it{AmazeFileManager\#1796} & 12 & 14.99 & \cmark & 17.03 & \cmark & 19.25 & \cmark & 21.49 & \cmark & 23.89 & \cmark & 51.46 & \cmark & 35.42 & \cmark & 18.28 & \cmark & \cellcolor{lightgray} 17.63 & \cellcolor{lightgray} \cmark \\
        
        \it{AmazeFileManager\#1837} & 4 & 3.70 & \cmark & 4.21 & \cmark & 4.87 & \cmark & 5.40 & \cmark & 6.56 & \cmark & 10.41 & \cmark & 15.62 & \cmark & 6.08 & \cmark & \cellcolor{lightgray} 5.48 & \cellcolor{lightgray} \cmark \\
        
        \it{and-bible\#261} & 18 & 20.31 & \xmark & 23.63 & \xmark & 27.29 & \xmark & 30.50 & \xmark & 33.88 & \xmark & 84.24 & \cmark & 46.45 & \xmark & 39.12 & \cmark & \cellcolor{lightgray} 36.01 & \cellcolor{lightgray} \cmark \\
        
        \it{AnkiDroid\#4200} & 13 & 14.54 & \xmark & 16.90 & \cmark & 19.94 & \cmark & 22.29 & \cmark & 24.04 & \cmark & 28.91 & \cmark & 21.77 & \cmark & 17.83 & \cmark & \cellcolor{lightgray} 16.86 & \cellcolor{lightgray} \cmark \\
        
        \it{AnkiDroid\#4451} & 19 & 23.42 & \cmark & 26.77 & \cmark & 30.51 & \cmark & 34.08 & \cmark & 37.64 & \cmark & 39.52 & \cmark & 42.91 & \cmark & 31.88 & \cmark & \cellcolor{lightgray} 29.61 & \cellcolor{lightgray} \cmark \\
        
        \it{AnkiDroid\#5638} & 4 & 3.58 & \cmark & 4.26 & \cmark & 4.85 & \cmark & 5.41 & \cmark & 6.01 & \cmark & 12.20 & \cmark & 6.33 & \cmark & 4.97 & \cmark & \cellcolor{lightgray} 4.04 & \cellcolor{lightgray} \cmark \\
        
        \it{AnkiDroid\#5756} & 16 & 17.92 & \xmark & 21.04 & \xmark & 23.67 & \xmark & 26.67 & \cmark & 29.76 &  \cmark & 34.91 & \cmark & 23.37 & \cmark & 21.10 & \cmark & \cellcolor{lightgray} 20.73 & \cellcolor{lightgray} \cmark \\
        
        \it{AnkiDroid\#6145} & 24 & 38.20 & \cmark & 43.17 & \cmark & 48.08 & \cmark & 52.21 & \cmark & 56.94 & \cmark & 69.15 & \cmark & 46.01 & \cmark & 46.05 & \cmark & \cellcolor{lightgray} 39.77 & \cellcolor{lightgray} \cmark \\
        
        \it{APhotoManager\#116} & 3 & 2.39 & \xmark & 2.86 & \xmark & 3.22 & \xmark & 3.63 & \xmark & 4.07 & \cmark & 10.06 & \cmark & 4.44 & \cmark & 2.88 & \cmark & \cellcolor{lightgray} 2.38 & \cellcolor{lightgray} \cmark \\
        
        \it{collect\#3222} & 9 & 9.49 & \xmark & 11.26 & \xmark & 12.74 & \cmark & 14.40 & \cmark & 16.00 & \cmark & 18.34 & \cmark & 13.14 & \cmark & 11.85 & \cmark & \cellcolor{lightgray} 10.59 & \cellcolor{lightgray} \cmark \\
        
        \it{geohashdroid\#118} & 4 & 3.67 & \xmark & 4.18 & \xmark & 5.11 & \xmark & 5.40 & \xmark & 6.06 & \xmark & 12.49 & \cmark & 13.33 & \cmark & 10.72 & \cmark & \cellcolor{lightgray} 9.89 & \cellcolor{lightgray} \cmark \\
        
        \it{Omni-Notes\#745} & 11 & 13.82 & \cmark & 15.79 & \cmark & 18.37 & \cmark & 19.57 & \cmark & 21.88 & \cmark & 28.94 & \cmark & 23.42 & \cmark & 18.60 & \cmark & \cellcolor{lightgray} 17.36 & \cellcolor{lightgray} \cmark \\
        
        \it{open-event-attendee\#2198} & 5 & 5.19 & \xmark & 5.49 & \xmark & 6.30 & \cmark & 7.27 & \cmark & 8.09 & \cmark & 14.66 & \cmark & 8.12 & \cmark & 6.59 & \cmark & \cellcolor{lightgray} 6.33 & \cellcolor{lightgray} \cmark \\
        
        \it{openlauncher\#67} & 4 & 5.21 & \xmark & 5.80 & \cmark & 6.38 & \cmark & 7.01 & \cmark & 7.61 & \cmark & 8.95 & \cmark & 6.68 & \cmark & 5.76 & \cmark & \cellcolor{lightgray} 5.31 & \cellcolor{lightgray} \cmark \\
        
        \it{Scarlet-Notes\#114} & 23 & 27.92 & \xmark & 32.50 & \cmark & 37.06 & \cmark & 40.80 & \cmark & 45.87 & \cmark & 67.78 & \cmark & 54.13 & \cmark & 34.61 & \cmark & \cellcolor{lightgray} 30.10 & \cellcolor{lightgray} \cmark \\
        
        \it{WordPress\#10302} & 3 & 2.48 & \cmark & 2.87 & \cmark & 3.29 & \cmark & 3.67 & \cmark & 4.09 & \cmark & 10.25 & \cmark & 5.10 & \cmark & 2.77 & \cmark & \cellcolor{lightgray} 2.32 & \cellcolor{lightgray} \cmark \\
        
    \hline
    \hline
    
    \it{Average} & 10.6 & 12.97 & 44\% & 14.94 & 66\% & 16.97 & 77\% & 18.75 & 83\% & 20.79 & 89\% & 31.31 & 100\% & 22.89 & 94\% & 17.51 & 100\% & \cellcolor{lightgray} 15.93 & \cellcolor{lightgray} 100\% \\
    \hline
    \end{tabular}
\end{table*}

% 412.09

\subsection{RQ2: Performance of \tool}
% \chen{Need to tell why we do this experiment. By using existing bugs, we want to see if our model may affect the bug triggering capability, may also demonstrate the negative effects of partially-rendered GUI for bug detection.}
% \chen{As mentioned in the intro, do not say such kind of thing which is of no evidence or proof.}
% As the bugs triggered from partially rendered GUIs might not be encountered by real-world users. 
% \chen{Update this, and please double check!}
Although we demonstrated the performance of our model in discriminating the rendering state of a single given GUI in the last RQ, it is still unclear if our approach can work in real-world GUI testing.
Therefore, we used the existing developer-verified bugs to evaluate the ability of the \tool to help efficiently test the app without affecting the bug triggering capability.

\textbf{Experimental Setup.}
% \revise{As executing events on partially rendered GUIs may hinder the following app exploration, resulting in the missing of bugs} \chen{This sentence is not complete, BTW, it should be used in the reasons why throttle@200/400/600 cannot replay bugs.}
To answer RQ2, we collected 18 crash bugs from 12 Android apps with defects studied in previous works~\cite{su2021benchmarking}.
Each crash bug has a trace script to reproduce, which will constantly trigger the app crash.
We evaluated our experiments under the common frame rate 30 fps.
% To answer RQ2, we evaluated the ability of our \tool to efficiently \revise{replay} the valid bug scenarios in app testing, without affecting the bug triggering capability.
% To accomplish this, we collected 18 \revise{crash scripts} from 12 Android apps with defects studied in previous works~\cite{su2021benchmarking}.
% \revise{When replaying the testing script, the app should constantly crash at the end.
% As the frame rate may affect the efficiency measurement, we evaluated our experiments under the common frame rate 30 fps.}
% Note that we only selected these testing seeds because they can firmly trigger the bugs over multiple trials.

% The higher the ratio of uncovered bugs, or mutants, the more effective a test suite is said to be

\textbf{Metrics.}
To measure the performance of our approach, we employed two evaluation metrics, i.e., whether the method can successfully reproduce the crash bug (\textbf{R}), and the time it takes to trigger the bug (\textbf{T}).
The less time it takes, the more efficient the method can trigger the bugs.
%\revise{The higher the ratio of bug reproducibility, the more effective to find the bugs encountered by real-world users.}
% For the success rate, a higher value represents more effectiveness to detect the bugs.

\textbf{Baselines.}
We set up 6 throttling methods as our baselines to compare with our \tool.
\textit{Throttle@k} is the fixed interval of k milliseconds between events.
We set the throttle k to 200ms, 400ms, 600ms, 800ms, 1000ms, as these throttle intervals are empirically used in automated testing~\cite{patel2018effectiveness}.
\textit{Themis}~\cite{su2021benchmarking} is the benchmark method of our experimental testing dataset.
It adopts a widely-used Android testing framework UIAutomator~\cite{web:uiautomator}, which explicitly waits between events until all resources are acquired.
% \chen{How to explore apps? You know the trace to trigger bugs? Or use droitbot to explore the app automatically? Please clarify it.}

In addition, we also add two derivatives of our approach to demonstrate the impact of each component. 
%To demonstrate the advantage of the \tool, we set up 2 ablation studies as our baselines.
In Section~\ref{sec:phase1}, we utilized heuristic image processing-based methods to calculate the similarity of consecutive frames to automatically discriminate the GUI rendering state.
Therefore, we set up one baseline called \textit{Consecutive Frame} based on multiple screenshots to compare with our method based on one single GUI screenshot.
In addition, to demonstrate the strength of real-time GUI state streaming outlined in Section~\ref{sec:phase3}, we also conducted an ablation study, namely \textit{Asynchronous Streaming}.
Specifically, it leverages native ADB built-in function~\cite{web:adb}, for example, it first adopts \texttt{adb screencap} to capture the GUI screenshot to the device, and then \texttt{adb pull} to transmit to the local machine for GUI rendering state classification, asynchronously.

\begin{figure}
	\centering
	\includegraphics[width=0.75\linewidth]{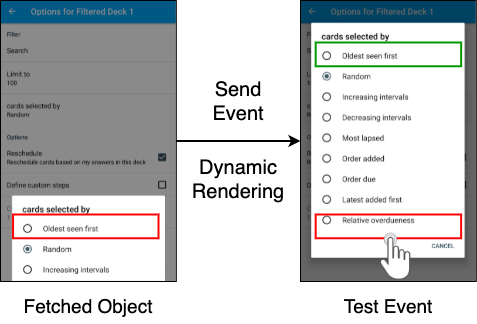}
	\caption{Example of failure bug detection with short throttle. Red box represents the test location, and green box represents the real object location.}
	\label{fig:rq2_example}
% 	\vspace{-0.5cm}
\end{figure}

\textbf{Results.}
Table~\ref{tab:rq2} shows detailed results of the time and reproducibility rate for each crash bug, where the number of steps to trigger the bug of each app is also displayed.
It takes \tool 15.93 seconds on average to reproduce all the bugs.
Instead, it takes the throttle methods of 200ms, 400ms, 600ms, 800ms, and 1000ms time intervals, on average 12.97, 14.94, 16.97, 18.75, and 20.79 seconds to reproduce the bugs, respectively.
The former three throttles only trigger fewer bugs than the latter two, especially the 200ms setting can only trigger 44\% of the bugs due to two reasons.
First, Fig.~\ref{fig:rq2_example} shows a failure example to trigger the bugs by using the short throttle (e.g., Throttle 200ms), that the event is executed on an out-of-sync object, due to the GUI is rendering, and the object position is dynamic.
Second, short throttle executing the events on partially rendered GUI can throw an unhandled exception bug.
However, these bugs will not be encountered by real-world users, instead, they hinder the following app exploration, resulting in the missing of the ``real'' bugs.
In contrast, our approach can efficiently trigger all of the crash bugs by discriminating partially rendered GUIs. %indicating the capability of \tool to detect bugs that are not causing by the partially rendering GUIs.

% \chen{May tell more example to show negative effects of partially rendered GUIs? I am even thinking about if we should say detecting partially-rendered GUI is one of our contributions?}

The benchmark method Themis can also trigger all of the bugs, but it takes much longer time, on average 31.31 seconds, which is 2x slower than our approach.
This is due to monitoring the potential resources can be time-consuming, including fetching GUI properties (e.g., widget type, location, and size) and explicitly waiting for lazy-loading assets (e.g., video). 
The more widgets in GUI, the longer it takes to determine the rendering state for Themis, as it monitors the rendering of all widgets in the GUI.
In contrast, we adopt a lightweight approach by leveraging easy-to-obtained GUI screenshots to adaptively adjust the throttle interval between events, obtaining efficiency without affecting the bug triggering capability.

Table~\ref{tab:rq2} shows the performance of ablation baselines of the \tool.
The heuristic image-processing method (Consecutive Frame) triggers 94\% of the crash bugs.
One failure case is that the resource loading is so slow beyond our threshold setting in Section~\ref{sec:phase1}, so it requires more frames to determine whether the GUI is fully rendered.
% \chen{Do not understand.}
In contrast, our approach can trigger all of the crash bugs in a shorter time, i.e., saving 30\% time on average.
This demonstrates the advantage of our approach of using a single GUI screenshot to discriminate the GUI rendering state, as multiple-screenshots capturing, transmitting, and computing take time.
In addition, leveraging a real-time GUI rendering monitor speeds up the testing process (9\% faster) than that of an asynchronous monitor (Asynchronous Streaming).
As a result, \tool does not affect the capability to trigger the bugs, especially those caused by partially rendered GUIs; on the other hand, \tool can speed up the automated testing, saving much of the time budget in hundreds or thousands of steps in long-term testing.

% We also manually observe the issues with our failure tests.
% Due to network latency, some states require longer intervals to load resources, which exceeds our maximum throttle limit ($>$ 1000ms).
% Therefore, our approach continues to send the next event, which executes on non-interactive GUIs (e.g., loading page).
% But we believe setting a longer maximum throttle threshold will potentially address the issue.

\subsection{RQ3: Usefulness of \tool}
\label{sec:rq3}
\textbf{Experimental Setup.}
To answer RQ3, we carried out a usefulness study to assess the performance of our \tool within the automated testing tools in the real-world testing environment.
To accomplish this, we utilized 32 Android apps which were used in our motivational study in Section~\ref{sec:motivation}.
They are top-rated on Google Play, covering 15 app categories.
% Details of these apps are shown in our online appendix\footnote{https://sites.google.com/view/fse2022}.

\textbf{Metrics.}
We employed three evaluation metrics to measure the performance of our \tool deployed in automated testing tools, i.e., activity coverage and GUIs.
% Activity coverage is a common metric used in automated testing evaluation, which measures the percentage of activities tested in the app.
For activity coverage, we collected all the activities defined in each app from AndroidManifest.xml following existing studies~\cite{cai2020fastbot,chen2019storydroid}, and measured the percentage of the explored activities for run-time. 
% A good activity coverage indicates that the method drives testing execution through a significant set of activity states at runtime.
Note that there may be multiple GUIs with different states in one activity, so we also used the GUIs for evaluation~\cite{li2017droidbot}.
The number of GUIs represents the number of events sent at run-time, and the number of fully-rendered GUIs represents whether the event is executed on a fully rendered GUI.
We also employed number of crashes to evaluate the ability of our tool in bug detection.
To ensure the crash validity, we verified them from app developers via issue tracker or direct contact.
% Note that we automatically replayed these crashes to filter the invalid crashes that would not be triggered if GUI is fully rendered.
For all metrics, a higher value represents better performance in automated app testing.
% Due to resource latency, performing actions to a partial rendering GUI may trigger some invalid bugs, which can be prevented when the GUI is fully rendered.
% Therefore, we manually measure the GUI quality to determine whether the tool is testing under valid GUI circumstances.
We recruited two paid annotators from online posting who have experience in GUI annotation, to annotate the quality of GUIs. 
To help ensure the validity and consistency of the annotation, we first asked them to spend twenty minutes distinguishing the difference between fully rendered GUIs and partially rendered GUIs.
Then, we assigned the set of captured GUI screenshots to them to annotate independently without any discussion. 
After the annotation, the annotators met and sanity corrected the subtle discrepancies. 
Any disagreement would be handed over to one author for the final decision. 
Note that the annotators and the author do not know whether the GUI is captured from our approach or baselines.

\textbf{Baselines.}
To demonstrate how our \tool can enhance real-world testing environments, we integrated our approach into the mature automated testing tool Droidbot~\cite{li2017droidbot}, namely Droidbot+\tool.
% \revise{We selected Droidbot as our experimental tool because of its open-sourced, well-documented, and widely-used as the infrustructure for other GUI testing works~\cite{liu2020owl,li2019humanoid}.}
In detail, Droidbot+\tool does not need to carefully set the throttle interval between events.
It is adaptive to GUI rendering, moving to the next event as soon as GUI rendering is complete.
We have released our integrated version of Droidbot to public\footnote{\url{https://github.com/sidongfeng/AdaT}}.
We set up 5 throttle intervals as our baselines, including 200ms, 400ms, 600ms, 800ms, and 1000ms.
Each method runs on the app for 10 minutes without interruption.
To ensure the validity of the comparison, we used the configurations in Droidbot, such as using greedy depth-first to search activities, randomly generating events, etc.

\begin{table}
    \centering
	\caption{Usefulness in real-world testing tool. ``FR'' denotes the fully rendered GUIs.}
	\label{tab:rq3}
	\begin{tabular}{l|c||c|c|c}
	\hline
	\multirow{2}{*}{\bf{Method}} & \bf{Throttle} & \multirow{2}{*}{\bf{Coverage}} & \multirow{2}{*}{\bf{\# Crashes}} & \bf{FR GUI}  \\
	 & \bf{(ms)} &  & & \bf{(Total)} \\
	\hline
	\multirow{5}{*}{Droidbot} & 200 & 36.19\% & 16 & 2,194 (2,832) \\
	 & 400 & 34.47\% & 15 & 1,758 (2,199) \\
	 & 600 & 35.97\% & 15 & 1,257 (1,501) \\
	 & 800 & 35.11\% & 14 & 750 (859) \\
	 & 1000 & 30.15\% & 9 & 462 (504) \\
	\hline
	Droidbot+\tool &  adaptive & \textbf{43.14}\% & \textbf{21} & \textbf{2,848 (3,207)} \\
	\hline
	\end{tabular}
\end{table}

\textbf{Results.}
Table~\ref{tab:rq3} shows the results of the performance between Droidbot and Droidbot+\tool.
Droidbot+\tool achieves a median activity coverage of 43.14\% across 32 Android apps, which is 6.95\% higher even compared with the best baseline (e.g., 36.19\% in Throttle 200ms).
This is because the dynamic wait in Droidbot+\tool allows access to more activities that short throttling might disrupt.
For example, when the GUI is in rendering progress, a short throttle testing might execute events on partially rendered GUIs to hinder exploration, or send backward events to abandon exploration.
In addition, Droidbot+\tool outperforms the baselines by exploring 3,207 GUI states, and 88.81\% are fully rendered. 
Overall, the results indicate the effectiveness and efficiency of \tool-enhanced tool in covering most of the activities, GUI states, and fully rendered GUIs, compared with the vanilla tool.
As more activities and states are explored, Droidbot+Ours triggers the most crash bugs (21) compared to the baselines.

\begin{comment}
Droidbot throttled at 1000ms can get 2.86\% more fully rendered GUIs compared with our approach due to two main reasons.
First, albeit the good performance of our model in discriminating fully rendered GUI, we still make wrong predictions.
Second, Droidbot throttled at 1000ms explores fewer events (504 vs 3,207) and tests fewer activities (30.15\% vs 43.14\%), where some unexplored GUIs may require longer intervals ($>$ 1000ms) to render, which will negatively affect the results.
Although Droidbot throttled at 1000ms can get higher percentge of fully rendered GUIs, it explores fewer events and fewer activities than our approach.
%\chen{I remember that the max interval set by our approach is also 1000ms, if so, why our approach achieves lower quality? Is that due to the wrong classification results in our model?}
%\sidong{Suppose 2000ms runtime, throttle at 1000ms can explore 2 states ($S_a$, $S_b$) and all fully rendered, resulting in 100\% fully rendered GUIs. For our approach, it takes 1000ms to explore $S_a$, $S_b$, and 1000ms to explore $S_c$. However, $S_c$ needs longer interval ($>$ 1000ms) to render, resulting in our approach can only achieve 66\% fully rendered GUIs.}
%As some states required longer intervals ($>$ 1000ms) are not explored, they achieve slightly more (2.86\%) fully rendered GUIs than ours.
Overall, the comparison of baselines indicates the effectiveness and efficiency of our \tool in covering most of the activities and fully rendered GUIs.
\end{comment}
\section{Threats to Validity}
In our experiments evaluating our model, threats to internal validity may arise from the leakage of testing dataset.
To mitigate this threat, we proposed the training-testing dataset split on apps, representing an unbiased testing set of GUIs.
Another potential confounding factor concerns the quality of dataset used to train, test, and evaluate our model performance.
% To help mitigate the first threat, we utilize the Rico dataset~\cite{?}, which have undergone several quality control mechanisms to ensure a diverse set of GUIs.
The use of automated mechanism to collect our dataset may generate some noise data.
To help mitigate the threat, we trained and evaluated our model in a large-scale dataset of 20,125 GUIs, that training a deep learning-based model with sufficient good data could tolerate a small amount of noise~\cite{sukhbaatar2014training,lecun2015deep}.

The main external threats to the validity of our work are the representative of the apps and the Android testing tools selected to evaluate the usefulness of our approach.
To mitigate this threat, we selected the 32 top apps from 15 different categories on the Google Play Store.
The selected apps vary greatly in their functionalities. 
% Details of these apps are shown in our online appendix~\footnote{?}.
To demonstrate the improvements that our approach can have on Android testing tools, we selected Droidbot as it is widely used in previous studies~\cite{liu2020owl,li2019humanoid}.
% First, Droidbot is a widely-used testing tools in previous studies~\cite{?}.
% Second, Droidbot is implemented in Python and is best suited for integration with deep learning model.
% While performing additional experiments with more testing tools is ideal, our implementation in Droidbot illustrate a reasonably usefulness of our tool, which should play a bigger role in other real-world practice.

Another threat to the validity arises from the randomness of the Android testing tools, apps, and emulators in our study and evaluation.
Namely, across different runs of the same tool, app, and emulator, the obtained metrics could change.
To mitigate this threat, we ran each pair of tools and apps two times, where each run was performed on a newly-created emulator with the same software and hardware configurations throughout all of the experiments.
The results were then from the aggregation of the two runs for each pair of tools and apps.

\section{Discussion}
% \chen{This section can be removed if more space is needed. It can be added into our future extension to the journal.}
\label{sec:discussion}
\textbf{Generality for automated testing tools.}
Results in RQ3 (Section~\ref{sec:rq3}) have initially demonstrated the usefulness of our approach in real-world practice when integrated into automated testing tools like Droidbot.
Our approach is a purely image-based method, relying only on GUI screenshots.
As the GUI screenshots are easy to capture in automated testing tools, our approach should play a bigger role in real-world practice.

\textbf{Generality across apps and platforms.}
Supporting tests on native and hybrid apps is a critical task in practice~\cite{lee2016hybridroid}.
As the GUI screenshots from different types of apps exert almost no difference, our approach can be generalized in testing different types of apps.
Another potential interest in automated testing is to support different platforms, e.g., iOS and Web.
While we focus on the Android platform for brevity in this study, the tool can be extended to other platforms.
We have conducted a small-scale experiment of 50 GUI screenshots from iOS and Web. Results show that our approach can achieve 92\% and 86\% F1-score in identifying GUI rendering state. We believe that the performance will be further boosted after fine-tuning.
We have released all of our model and source code for reproducibility.

\textbf{Collections of High-quality GUI dataset.}
A large-scale of GUI collection is the foundation for many downstream deep-learning based GUI related research such as code generation~\cite{chen2018ui,chen2019storydroid,chen2019gui,moran2018machine,feng2021auto,feng2022auto}, GUI design~\cite{chen2019gallery,chen2020wireframe,chen2020lost,reiss2018seeking,zhao2019actionnet,feng2022gallery,chen2022automatically}, GUI testing~\cite{moran2018automated,chen2020object,su2022metamorphosis,liu2022nighthawk,zhao2020seenomaly,yang2021don,xie2020uied,xie2022psychologically}, etc.
Existing studies~\cite{li2020mapping} have identified the limitations in the mobile GUI dataset, and attempted to denoise the dataset based on GUI widget classname and GUI layout.
%In contrast to filtering the noise data (e.g., partially rendered GUI)
Our work complement with noise removal study, as results in RQ3 (Section~\ref{sec:rq3}) have illustrated the benefit of our approach in collecting high-quality (e.g., fully rendered) GUIs, laying a solid foundation for other works in this direction.

% \textbf{Future work.}

\section{Related Work}
As our work is to utilize GUI rendering state to tackle the efficiency issue for accelerating automated GUI testing, we introduce related works in two aspects, i.e., automated GUI testing, and efficiency support for testing.

\subsection{Automated GUI Testing}
A growing body of tools has been dedicated to assisting in automated app testing.
One of the earliest efforts is Monkey~\cite{web:monkey}, Google’s official testing tool for Android apps, intended for generating random user events such as clicks, touches, or gestures, as well as a number of system-level events on the GUI.
Subsequent efforts have led to test case generation based on randomness strategies~\cite{mao2016sapienz,machiry2013dynodroid,ye2013droidfuzzer}, or app artefacts (e.g., activity, source code)~\cite{azim2013targeted,yang2013grey}.
% One of the most efficient approaches is the random generation of event sequences~\cite{?}.
% Monkey~\cite{?} is Google’s official testing tool for Android apps, which is built into the Android platforms and widely used by developers.
% Monkey generates random user events such as clicks, touches, or gestures, as well as a number of system-level events on the GUI.
% Mao et al.~\cite{?} developed a multi-objective automated testing technique Sapienz, leveraging genetic algorithms to optimize randomly generated tests to maximize code coverage while minimizing event sequences.
% Another popular approach of Android app testing is based on app artefacts, such as activities, source code, and configuration files. 
% For example, DroidBot~\cite{?} uses different methods to dynamically construct an activity transition graph on-the-fly and consume it to generate test events. 
% A3E~\cite{?} utilizes a hybrid method of depth-first exploration and breadth-first exploration to systematically explore the activities in the apps.
% ORBIT~\cite{?} is a white-box approach that adopts the source code of the app to determine the events that can be tested.

Recent tools~\cite{liu2022guided,li2017droidbot,amalfitano2012using} leverage dynamic and static analysis to reverse engineer a stochastic model from GUI to generate more robust automated testing.
Gu et al.~\cite{gu2019practical} present a GUI event-refinement model, that uses GUI runtime information to evolve an initial model to generate precise events.
Su et al.~\cite{su2017guided} propose Stoat that assigns GUI runtime widgets with different probabilities of being selected to achieve effective testing. 
Moreover, computer vision techniques have been applied to further improve the effectiveness of automated GUI testing.
Degott et al.~\cite{degott2019learning} adopt reinforcement learning to identify valid interactions for a GUI element (e.g., a button allows to be clicked but not dragged) to guide testing.
Li et al.~\cite{li2019humanoid} take a sequence of GUIs captured from manual events to learn a model to predict human-like interactions on the given app.
Different from these approaches that focus on sophisticated GUI algorithms for achieving higher test coverage, our approach aims to accelerate automated testing by scheduling the test events with GUI rendering status inference, leading to substantial testing efficiency and effectiveness improvement.

\subsection{Efficiency Support for Testing}
There have been many works trying to improve infrastructure support for the purpose of efficient testing.
Hu et al.~\cite{hu2014efficiently} propose AppDoctor that instruments the target apps using invocations of event handlers to quickly find potential sequences of error-triggering GUIs.
Song et al.~\cite{song2017ehbdroid} improve the efficiency of AppDoctor by leveraging direct invocations. 
Wang et al.~\cite{wang2021infrastructure} propose an Android tool Toller that injects into the testing device to efficiently access GUI layout and execute events.
Different from those infrastructure support, we aim to speed up automated testing by adaptive throttling, scheduling the testing events for efficiency improvement.

Adaptive throttling is a common practice for efficient testing on the web.
Selenium~\cite{web:selenuim} proposes a feature called Explicit Wait to tell the testing driver to wait for an explicit amount of time until the presence of elements.
Similar to Selenium, many tools build this feature for mobile testing, such as Appium~\cite{web:appium}, UIAutomator~\cite{web:uiautomator}, etc.
Specifically, those tools verify the presence of elements by fetching the view hierarchy of the GUI.
However, subsequent studies~\cite{li2020mapping} find that the fetched GUI views may be out of sync, leading to tests on misaligned or invalid objects.
Furthermore, those tools only check the validity of the GUI view hierarchy, while not resources, which may restrict the capability of testing exploration.
In contrast, we leverage the GUI as a whole with visual information to dynamically adjust the throttle to schedule the events when the GUI is fully rendered, which is analogous to human viewing and interacting.

\section{Conclusion}
Automated app testing is crucial to improve app quality.
Despite the numerous automated testing tools, one often overlooked aspect is the throttle between events.
A short throttle may reduce the effectiveness of testing, while a long throttle may reduce the efficiency of testing.
To strike the balance, we propose \tool, a lightweight image-based approach to adaptively adjust the throttle based on the GUI rendering inference.
Given the real-time streaming on GUI, \tool adopts a deep learning model to infer the rendering state to adjust events scheduling, sending events when the GUI is fully rendered.
The experiments demonstrate the performance and usefulness of our approach in improving the efficiency and effectiveness of automated testing.

In the future, we will keep improving our \tool for better efficiency in two aspects.
First, we can reduce the computational cost of inference by optimizing network architecture.
Second, the process of our \tool can be accelerated by more advanced infrastructure support.
We also want to deploy our approach to mobile devices, so that it can be applied to on-device testing.
It can be achieved by quantizating our model into a lite-based model.

\section{Acknowledgement}
This work is partially supported by the Monash FIT RSP fund.

\bibliographystyle{IEEEtran}
	\bibliography{main}
\end{document}